\documentclass{vgtc}                          % final (conference style)
\graphicspath{{figures/}{pictures/}{images/}{./}} % where to search for the images

\usepackage{times}                     % we use Times as the main font
\usepackage{tabu}                      % only used for the table example
\usepackage{booktabs}                  % only used for the table example
\usepackage{lipsum}                    % used to generate placeholder text
\usepackage{mwe}                       % used to generate placeholder figures

\newif\ifshowfeedback
\showfeedbackfalse

\usepackage{subcaption}
\usepackage{multirow} % Add this in your preamble
\usepackage{siunitx}
\usepackage{booktabs}
\usepackage{amsfonts}
\usepackage{booktabs}
\usepackage{siunitx}
\usepackage{makecell}
\usepackage{arydshln}
\usepackage{array}
\usepackage{soul}
\usepackage[linewidth=1pt]{mdframed}
\usepackage{tikz}
\usepackage{amsmath}
\usepackage{float}
\usepackage{layouts}
\usepackage{amsmath}
\usepackage{algorithm}
\usepackage{algpseudocode}
\usepackage{comment}
\usepackage{outlines}
\usepackage{xcolor,cancel}
\usepackage{xcolor}
\usepackage{diagbox}
\usepackage{amssymb}% http://ctan.org/pkg/amssymb
\usepackage{pifont}% http://ctan.org/pkg/pifont
\usepackage[most]{tcolorbox}
\tcbset{
  colback=gray!10,
  colframe=black,
  boxrule=0.3mm,
  arc=1mm,
  left=3mm,
  right=3mm,
  top=1mm,
  bottom=1mm,
  boxsep=1mm,
  fonttitle=\bfseries,
  breakable,
  enhanced,
  before upper=\strut}

\makeatletter
\@ifundefined{ifshowfeedback}{
  \newif\ifshowfeedback
  \showfeedbacktrue   % or \showfeedbackfalse
}{}
\makeatother

\ifshowfeedback
  \newcommand{\feedbackrequest}[1]{%
    \textbf{\textcolor{blue}{\small\underline{FEEDBACK REQUEST:}~#1}}\par}
\else
  \newcommand{\feedbackrequest}[1]{}
\fi

\usepackage[normalem]{ulem}
\usepackage{amsmath}
\newcommand{\stkout}[1]{%
  \ifmmode
    \text{\textcolor{red}{\sout{\ensuremath{\textcolor{black}{#1}}}}}%
  \else
    \textcolor{red}{\sout{#1}}%
  \fi
}

\usepackage{enumitem}
\usepackage{soul,xcolor}

\definecolor{darkgreen}{rgb}{0.13, 0.55, 0.13}
\ifshowfeedback
    \setstcolor{red}
  \newcommand{\review}[2]{%
    {\st{#1}}~\textcolor{darkgreen}{\textbf{#2}}%
  }
\else
  \newcommand{\review}[2]{#2}
\fi

\ifshowfeedback
  \newcommand{\ejainroundone}[1]{%
    \textbf{\textcolor{red}{\small\underline{EJ:}~#1}}\par}
\else
\newcommand{\ejainroundone}[1]{}
\fi

\ifshowfeedback
  \newcommand{\aibragimovroundone}[1]{%
    \textbf{\textcolor{purple}{\small\underline{AI:}~#1}}\par}
\else
\newcommand{\aibragimovroundone}[1]{}
\fi

\ifshowfeedback
  \newcommand{\ethanwilsonroundone}[1]{%
    \textbf{\textcolor{orange}{\small\underline{EW:}~#1}}\par}
\else
\newcommand{\ethanwilsonroundone}[1]{}
\fi

\ifshowfeedback
  \newcommand{\michaelproulxroundone}[1]{%
    \textbf{\textcolor{teal}{\small\underline{MP:}~#1}}\par}

\else
\newcommand{\michaelproulxroundone}[1]{}
\fi

\usepackage{mathptmx}                  % use matching math font

\onlineid{0}

\vgtccategory{Research}

\vgtcinsertpkg

\title{Toward Multimodal Privacy in XR: Design and Evaluation of Composite Privatization Methods for Gaze and Body Tracking Data}

\author{Azim Ibragimov\thanks{e-mail: a.ibragimov@ufl.edu}\\ %
        \scriptsize University of Florida %
\and Ethan Wilson\\ %
     \scriptsize University of Florida %
\and Kevin R. B. Butler\\ %
     \scriptsize University of Florida %
\and Eakta Jain \\ %
    \scriptsize University of Florida }

\abstract{
As extended reality (XR) systems become increasingly immersive and sensor-rich, they enable the collection of behavioral signals such as eye and body telemetry. These signals support personalized and responsive experiences and may also contain unique patterns that can be linked back to individuals. However, privacy mechanisms that naively pair unimodal mechanisms (e.g., independently apply privacy mechanisms for eye and body privatization) are often ineffective at preventing re-identification in practice. In this work, we systematically evaluate real-time privacy mechanisms for XR, both individually and in pair, across eye and body modalities. We assess privacy through re-identification rates and evaluate utility using numerical performance thresholds derived from existing literature to ensure real-time interaction requirements are met. We evaluated four eye and ten body mechanisms across multiple datasets, comprising up to 407 participants. Our results show that when carefully paired, multimodal mechanisms reduce re-identification rate from 80.3\% to 26.3\% in casual XR applications (e.g., VRChat and Job Simulator) and from 84.8\% to 26.1\% in competitive XR applications (e.g., Beat Saber and Synth Riders), all while maintaining acceptable performance based on established thresholds. To facilitate adoption, we additionally release XR Privacy SDK, an open-source toolkit enabling developers to integrate the privacy mechanisms into XR applications for real-time use. These findings underscore the potential of modality-specific and context-aware privacy strategies for protecting behavioral data in XR environments.
} % end of abstract


\begin{document}


\firstsection{Introduction}

\maketitle

Extended reality (XR) systems, including virtual reality (VR), augmented reality (AR), and related immersive technologies, are rapidly growing in adoption across consumer, enterprise, and research sectors~\cite{lee_adoption_2019, chuah2019wearable}. To deliver immersive and interactive experiences, XR headsets collect fine-grained behavioral telemetry through multiple onboard sensors, including eye and body telemetry. 

% Among these, eye telemetry captures time-series data on gaze direction and movements, recorded via eye-tracking sensors. 
% It supports interaction~\cite{monteiro2021hands, fernandes2023leveling, atienza2016interaction}, avatar animation~\cite{normoyle2013evaluating, ruhland2015perception, seele2017here}, attention analysis~\cite{liu2022impacts, wang2021prior}, rendering optimization~\cite{kaplanyan2019deepfovea, david2021towards}, and biometric authentication~\cite{hallal2024recent, bozkir2023eye, lohr2022Eyeb, lohr2024Establishing}. 
% Body telemetry is a time series recorded via the headset and controllers, capturing head and hand movements as well as static attributes like height and wingspan.
% It is used for interaction~\cite{yang2019gesture, cabral2005usability}, avatar control~\cite{ji2021use}, and behavioral biometrics~\cite{miller2020personal, jones2021literature}. These sensing modalities form the basis for most real-time interaction and personalization in XR.

Eye telemetry captures time-series data on gaze direction and movements via eye-tracking sensors~\cite{monteiro2021hands, fernandes2023leveling, atienza2016interaction}, while body telemetry, recorded via the headset and controllers, tracks head and hand movements along with static attributes such as height and wingspan~\cite{yang2019gesture, cabral2005usability, ji2021use}. Together, these modalities enable essential control functions in XR: gaze tracking supports selection and targeting~\cite{monteiro2021hands, fernandes2023leveling, atienza2016interaction}, and body tracking through head and controller motion allows navigation, object manipulation, and physical gameplay~\cite{yang2019gesture, cabral2005usability}. Unlike modalities such as voice or facial expression, which can be disabled without affecting most experiences, eye and body signals are always active and indispensable. Disabling them would remove core movement and interaction capabilities, making most XR applications unusable. This non-optional nature makes them uniquely challenging from a privacy perspective.

% \begin{itemize}
% \item \textbf{Eye telemetry:} Time-series data capturing gaze direction and eye movement dynamics. Used for interaction~\cite{monteiro2021hands, fernandes2023leveling, atienza2016interaction}, animation~\cite{normoyle2013evaluating, ruhland2015perception, seele2017here}, attention analysis~\cite{liu2022impacts, wang2021prior}, rendering optimization~\cite{kaplanyan2019deepfovea, david2021towards}, and biometric authentication~\cite{hallal2024recent, bozkir2023eye, lohr2022Eyeb, lohr2024Establishing}.
% \item \textbf{Body telemetry:} Time-series sensor data capturing head and hand motion in XR. Captured through head and hand tracking to enable interaction~\cite{yang2019gesture, cabral2005usability}, avatar animation~\cite{ji2021use}, and behavioral biometrics~\cite{miller2020personal, jones2021literature}.
% \end{itemize}

% While these behavioral signals enable highly personalized and responsive XR experiences, they also create privacy vulnerabilities. 
Research has shown that users can be re-identified based on eye movements~\cite{lohr2020metric, lohr2022Eyeb}, body movements~\cite{miller2020personal, nair2023Uniquea}, or a combination of both~\cite{tricomi2209you, olade2020BioMove, jarin2023behavr}. In this context, re-identification refers to the ability of a model or adversary to correctly match a sample of behavioral data (e.g., eye or body movements) to the person from whom it was collected. 
% Beyond re-identification, behavioral data can enable inference of sensitive attributes such as gender~\cite{ tran2022comparing, giaretta2024security, 10582036}, age~\cite{zhang2018old, tran2022comparing, giaretta2024security}, cognitive conditions~\cite{armstrong2012eye, giaretta2024security}, and sexual orientation~\cite{renaud2002measuring, giaretta2024security}. 
% Unlike traditional identifiers (e.g., usernames), behavioral signals are difficult for users to consciously control or suppress, creating unique privacy challenges.

To protect user privacy, prior work has proposed privacy mechanisms for perturbing eye~\cite{wilson2024PrivacyPreserving, david-john2021Privacypreservingc, li2021kalvarepsilonido} or body~\cite{nair2023deep, nair2023Going} movements to reduce re-identification risk. 
% However, most existing approaches address each modality separately and are insufficient in real-world multimodal settings, where adversaries can leverage unprotected signals from another modality  for re-identification attacks~\cite{aziz2024exploiting}.
In multimodal settings, however, the challenge is greater. Recent work by Aziz et al.~\cite{aziz2024exploiting} provided an important first step by showing that even when protections are applied independently to both eye and body modalities, some residual identity information can persist, enabling cross-modal re-identification. Their study paired one eye privacy mechanism with one body privacy mechanism, illustrating that certain combinations offer only partial protection and highlighting the need for broader exploration of pairing strategies. They also emphasized the need for more rigorous evaluations, including testing across additional datasets, a wider variety of privacy mechanisms, the inclusion of utility assessments for future work and evaluation on attribute attacks. Building on this foundation, our work evaluates a broader set of mechanisms and usage contexts, examines how protections interact with application-specific usability requirements, and makes these methods accessible through a publicly available open-source SDK to facilitate real-time adoption in XR applications.

Beyond effectiveness alone, we also raise the question of how application context shapes these requirements. Different XR application types may impose distinct usability constraints, making some privacy mechanisms acceptable in one setting but disruptive in another. User performance is generally maintained when body telemetry deviations are within 21 cm Euclidean distance \cite{wentzel2020improving} and eye telemetry errors are under 2 degrees angular deviation \cite{schuetz2020psychophysics}, but these thresholds may play out differently depending on whether the experience prioritizes spatial precision or reaction time. For example, casual applications, such as Job Simulator~\cite{jobsimulator} and VRChat~\cite{vrchat}, often involve tasks like object manipulation and virtual browsing, which prioritize spatial accuracy over temporal responsiveness. In these scenarios, small delays introduced by privacy mechanisms may be acceptable if interaction precision is preserved. In contrast, competitive applications, such as Beat Saber~\cite{beatsaber} and Synth Riders~\cite{synthriders}, demand low latency and fast response times, where even minor delays can significantly degrade user performance.
Thus, any privacy protection must balance privacy gains against the real-time interaction requirements of different XR applications.

% \begin{itemize}
% \item \textbf{Casual applications:} Experiences such as object manipulation and virtual browsing prioritize spatial accuracy over speed. Small temporal delays introduced by privacy mechanisms may be acceptable if interaction precision is preserved.

% % \item \textbf{Competitive applications:} High-intensity applications such as Beat Saber~\cite{beatsaber}, where low-latency and fast response times are critical for task success. Even minor delays can substantially degrade user performance. 
% \end{itemize}

In this work, we systematically evaluate real-time multimodal privacy mechanisms for XR, tuned to maintain usability based on empirically grounded thresholds. We are the first to introduce and evaluate the role of application context, showing that the utility of privacy mechanisms differs between casual and competitive XR use cases.
Our results show that 
combining privacy mechanisms across modalities must be done carefully: not all  {pairs} improve privacy, and poorly  {paired mechanisms}  may leave residual vulnerabilities. We demonstrate that  {well-paired} multimodal perturbations can reduce re-identification risks from 80.3\% to 26.3\% for casual applications and from 84.8\% to 26.1\% for competitive applications, while still satisfying the real-time usability requirements specific to each application context.

% \textbf{In summary, our work differs from prior multimodal XR privacy research by:
% } 
\textbf{Contributions.}
Our work advances multimodal XR privacy research beyond prior studies~\cite{aziz2024exploiting} by:
\begin{enumerate}
    \item Presenting a large-scale evaluation of 14 real-time privacy mechanisms, including 4 gaze telemetry and 10 body telemetry, conducted on XR datasets with up to 407 participants.
    \item Proposing a context-aware evaluation framework that supports mechanism selection under usability constraints and derive design guidelines tailored to casual and competitive XR applications.
    \item Proposing and evaluating multimodal mechanism pairings that reduce re-identification from 80.3\% to 26.3\% in casual XR applications and from 84.8\% to 26.1\% in competitive XR applications, without violating usability thresholds.
    \item \textcolor{black}{Developing \texttt{XR-Privacy SDK}, a Unity-based toolkit that allows developers to incorporate privacy mechanisms into XR applications, supporting real-time protection of user data.}
\end{enumerate}

The rest of the paper is structured as follows. Section 2 reviews related work on identity inference, re-identification risks, and privacy mechanisms in XR, covering eye, body, and multimodal data streams. Section 3 presents the system overview and adversary models assumed in this work. Section 4 analyzes the privacy–utility tradeoffs for each modality in casual XR applications, while Section 5 does the same for competitive applications. Section 6 selects operating points within these tradeoffs based on real-time constraints and evaluates multimodal re-identification performance under those constraints. Section 7 presents XR-Privacy SDK, while Section 8 discusses future work and concludes the paper.

\section{Background}

Re-identification risks have been extensively considered in the related literature. Prior work has also proposed various mitigation strategies to address these privacy concerns. In this section, we review the literature on re-identification and practical privacy mechanisms designed to reduce re-identification risk.

\subsection{Re-Identification Risks from Recorded Eye  Telemetry}

Eye telemetry biometrics can allow user identification within human-computer interaction systems based on unique gaze behaviors~\cite{kasprowski2004eye}. Key features include fixations, periods where the gaze is stationary, and saccades, rapid transitions between fixation points, both of which exhibit user-specific patterns. In the past decade, eye-based re-identification has progressed from hand-crafted feature approaches to deep learning methods for end-to-end modeling~\cite{hallal2024recent, bozkir2023eye, lohr2022Eyeb}. The current state-of-the-art, EyeKnowYouToo (EKYT), achieves 67.31\% accuracy on VR-based eye tracking~\cite{lohr2023GazeBaseVR, wilson2024PrivacyPreserving}.

\ethanwilsonroundone{Because re-id rates are completely unstandardized, you should also list EKYT's EER performance (or outright replace the re-id metrics with EER).}
\aibragimovroundone{I think there is still value in providing both Rank-1 accuracy and EER. EER is useful because it is standardized, and Rank-1 because it aligns with the metrics we use in our results. So, I plan to include both. Let me know if this works!
}

\review{}{Despite their utility, eye movement biometrics introduce potential privacy considerations. Gaze patterns are difficult to consciously alter and may serve as stable identifiers across sessions, devices, and applications. In immersive environments, eye tracking often operates passively in the background, raising the possibility of re-identification without explicit user awareness~\cite{wilson2024PrivacyPreserving}. Even when traditional identifiers (e.g., usernames) are removed, behavioral signals in gaze data could still allow linkage over time. These concerns may be more pronounced for underrepresented groups, who could be more easily distinguished in datasets with marginalized populations, who might face disproportionate harm if privacy is compromised ~\cite{sannon2022privacy}}.

\ethanwilsonroundone{This section header is "potential risks" but risks aren't talked about.  Only re-id performance is listed.}
\michaelproulxroundone{Also, some of the work here is focused on the 'benefits' of eye telemetry for authentication rather than the risks per se. It seems that balancing the pros/cons and costs/benefits would be better here.}
\aibragimovroundone{I agree. I added a paragraph saying that re-identification has risks associated with unwanted re-identification and that it is especially concerning for marginalized populations. Let me know if that works}

% Eye movement re-identification raises substantial privacy concerns, as individuals cannot easily obscure their gaze behavior during typical device usage. To address these risks, several privacy-preserving mechanisms have been proposed in the literature, including Gaussian noise addition, temporal downsampling, spatial quantization, and, most recently, smoothing~\cite{david-john2021Privacypreservingc, wilson2024PrivacyPreserving}. These techniques operate in real time and aim to balance privacy protection with interaction fidelity in XR applications.

\subsection{Re-Identification Risks from Recorded Body  {Telemetry}}

Body  {related} biometrics, particularly gait-based identification, have been extensively studied in the fields of computer vision and behavioral biometrics~\cite{lin2022gaitgl, chao2019gaitset, balazia2018gait, barnachon2014ongoing, patrona2018motion}. \review{More recently, this modality has been applied to XR systems, where  body telemetry data is continuously collected from head-mounted displays and hand-held controllers}{Traditionally, this line of research relied on high-fidelity motion capture systems that track full-body skeletal data with dozens of joint points~\cite{balazia2018gait, barnachon2014ongoing, patrona2018motion}. However, recent studies in XR settings demonstrate that even sparse body telemetry, typically limited to the positions and orientations of the head-mounted display and handheld controllers, can be surprisingly distinctive~\cite{hallal2024recent, giaretta2024security, mustafa2018unsure, rogers2015Approach, li2016whose}.}
\ethanwilsonroundone{Can you clarify here that the majority of research in this space has looked at higher fidelity animation skeletons (from motion capture usually), yet recent work in XR is successful with 3 points (head + controllers)?}
\aibragimovroundone{Agree - let me know if this works!}
% Early approaches relied on head  {telemetry} data, such as rotational velocity and translational acceleration, for user re-identification~\cite{mustafa2018unsure, rogers2015Approach, li2016whose}. As XR systems began incorporating motion-tracked controllers, including head and hand positions and orientations, became increasingly informative for identity inference~\cite{pfeuffer2019Behaviourala, nair2023Uniquea, miller2020personal}. 
The highest reported performance to date is achieved by an LSTM-based model, which attains up to 96.5\% re-identification accuracy using  {position and orientation} data from both the headset and controllers~\cite{nair2023deep}.
%\review{}{
% While there are legitimate uses for body telemetry recognition in XR—such as enabling seamless authentication or personalized interaction—it may also encode behavioral patterns that are difficult for users to consciously suppress. 

\review{}{
Similar to eye tracking data, which passively reflects unique behavioral traits, body movement patterns differ from traditional identifiers such as usernames in that they emerge naturally through interaction and are often produced without the user’s awareness. Since XR systems capture this telemetry continuously, tracking how individuals move their head and hands during everyday use, such data can become a consistent behavioral signal. In contexts where telemetry is logged or shared across sessions or applications, there exists a potential for this signal to be linked back to the same individual over time.
}{}
\ethanwilsonroundone{This section as well.  The header mentions risks from body movements, but the section only talks about identification performance.}
\aibragimovroundone{done - let me know if this works}

\subsection{Privacy Considerations in XR Applications}

To address the extraction of biometric features without user consent, prior work has proposed several privacy-preserving mechanisms aimed at reducing the risk of unwanted re-identification.

\ethanwilsonroundone{It's conflicting to say "a significant amount of research", then to only list one paper.  Kaleido is probably worth a mention here (just mention that it introduces latency).}

\ethanwilsonroundone{The methods are not detailed in Section 5.2.}
\aibragimovroundone{Thanks for pointing out the error in sectino number! And also I incorporate a discussion on other mechanisms, including Kaleido!}

 \textbf{Eye Movement Privacy Mechanisms.}
Significant research has focused on real-time protections for eye-tracking data. Wilson et al. evaluated multiple techniques, including Gaussian noise addition, spatial downsampling, temporal downsampling, and smoothing~\cite{wilson2024PrivacyPreserving}. These methods are detailed in Section \review{5.2}{4.2}. A key consideration is their impact on user performance. To that end, a user study was conducted to evaluate the trade-offs across techniques and parameter settings. Results indicate that smoothing  {privacy mechanism} offers the most favorable balance between privacy and usability. Other examples of privacy mechanisms include formal privacy guarantees such as differential privacy (DP)~\cite{liu2019Differentialb, steil2019privacy}, as well as frameworks like k-anonymity and plausible deniability~\cite{david-john2022Youra, david-john2023Privacypreserving}. However, as explained by Wilson et al.\cite{wilson2024PrivacyPreserving}, these algorithms are not suitable for real-time use. Kal$\epsilon$ido, for instance, employs a sample-level method with DP guarantees\cite{li2021kalvarepsilonido}, but it introduces significant computational overhead. Every 2.3 seconds, the mechanism runs an ROI detection algorithm that adds 80 ms of latency~\cite{li2021kalvarepsilonido}. As Wilson et al. note, this overhead is particularly problematic in XR applications, where system latency beyond 15–20 ms can cause discomfort or motion sickness~\cite{wilson2024PrivacyPreserving}. The fact that Kal$\epsilon$ido incurs this penalty every 2.3 seconds makes it impractical for such settings.

 \textbf{Body Movement Privacy Mechanisms.}
Real-time privacy mechanisms have also been developed for body  {telemetry} data, with a focus on minimizing utility degradation. Two notable approaches, Body-LDP~\review{}{~\cite{nair2023Going}} and Deep Motion Masking (DMM)~\review{}{~\cite{nair2023deep}}, are described in Section ~\review{5.2.}{4.2.} Nair et al. conducted empirical evaluations to measure their privacy-utility trade-offs  {for competitive applications}, finding that DMM provides stronger privacy guarantees with least impact on user interaction~\review{}{~\cite{nair2023deep}}.

\begin{figure*}
\centering
\includegraphics[width=0.85\textwidth]{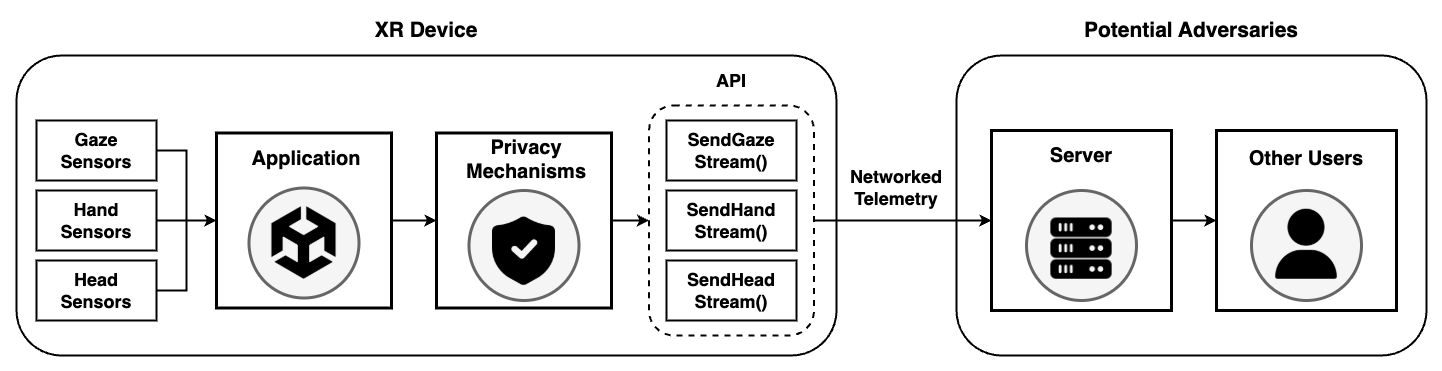}
\caption{\textbf{System overview.} Behavioral signals from gaze, head, and hand sensors are accessed by XR applications and transmitted via modality-specific API functions (e.g., \texttt{SendGazeStream()}) as network telemetry. These streams may be received by external servers or users, where adversaries can perform re-identification attacks.}
\label{fig:threat_model}
\end{figure*}

 \textbf{Multimodal Privacy Mechanisms.}
Multimodal privacy in XR remains an emerging area. Recent work by Aziz et al.~\cite{aziz2024exploiting} demonstrates that adversaries can bypass unimodal protections by exploiting unprivatized modalities. Moreover, even when both gaze and body signals are independently privatized, residual identity information may persist, enabling successful re-identification. \textcolor{black}{However, as the authors themselves acknowledge, their evaluation focused on a limited set of datasets and privacy mechanisms. In addition, the study did not incorporate evaluations to assess the usability of the proposed protections. Our work picks up on these two limitations and proposes multimodal evaluations on three datasets, fourteen privacy mechanisms, provides utility evaluations, and releasing the open-source XR-Privacy SDK, a Unity-based toolkit that enables XR developers to integrate privacy mechanisms into their own custom XR applications.}

\section{System and Threat Assumptions}
\subsection{System Overview}

We consider the risk of unauthorized user re-identification in networked XR systems. As illustrated in Figure~\ref{fig:threat_model}, XR headsets and controllers continuously capture behavioral telemetry, such as gaze direction, hand position, and head orientation, via on-board sensors. This data is exposed to applications through structured APIs (e.g., \texttt{SendGazeStream()}, \texttt{SendHandStream()}, \texttt{SendHeadStream()}), and transmitted to remote servers for logging, analytics, or multi-user  {experiences}.

We assume a potential adversary with legitimate access to these APIs, such as third-party applications or peer users, who operate within the scope of granted permissions. The adversary does not require a system compromise or unauthorized access. Instead, they analyze telemetry to infer user identity, either in real time or offline. 
% When only one modality is accessible, the potential adversary exploits unimodal features; when both eye and body data are available, \review{cross-modal correlations may be leveraged to bypass unimodal protections.}{ the adversary can use multimodal identification methods.}

 We assume privacy mechanisms are deployed within the XR device. These mechanisms operate after the application has processed XR telemetry, but before the data is exposed via API calls to external entities.  By introducing perturbations to eye and body movement data, the mechanisms reduce re-identification risk while negatively impacting utility, reflecting a privacy–utility trade-off.

 \subsection{Threat Model}
We adopt a threat model consistent with prior work on multimodal XR privacy~\cite{aziz2024exploiting}. The adversary is assumed to have access to XR telemetry collected from user sessions and attempts to infer user identities based on these behavioral signals. The attacker may be unaware of which, if any, privacy mechanisms have been applied to the data streams. However, through repeated observations and experimentation, they may infer which modalities are unprotected and adapt their strategy accordingly.

We evaluate three deployment scenarios: (a) no privacy mechanisms are applied, (b) all modalities are protected, and (c) only a subset of modalities are protected. This model captures a realistic adversary capable of exploiting cross-modal leakage, and enables us to assess the robustness of unimodal and multimodal defenses under varying levels of protection.

% \subsection{Research Questions}

% Given the system setup and adversarial capabilities described above, our analysis is guided by the following research questions:

% \begin{enumerate}[label=\textbf{RQ\arabic*:}, leftmargin=2.85em]
%     \item Among eye and body privacy mechanisms, which provide the best trade-off between re-identification risk and interaction fidelity in casual XR applications?
%     \item In competitive XR applications, which eye and body privacy mechanisms most effectively balance re-identification protection with the competitive interaction demands required for high-performance user experiences?
%     \item When streaming both eye and body telemetry data, how well do paired multimodal privacy mechanisms protect against re-identification in casual and competitive XR applications, while maintaining real-time usability?
% \end{enumerate}

\section{Evaluation of Anonymization in casual XR applications}

\subsection{Privacy Mechanisms}
\renewcommand{\arraystretch}{1}{
\begin{table*}[t]
\centering
\small
\caption{Summary of privacy mechanisms evaluated in this work.}
\begin{tabular}{!{\vrule width 0.8pt}p{1.7cm}!{\vrule width 0.8pt}p{8.0cm}!{\vrule width 0.8pt}p{3.5cm}!{\vrule width 0.8pt}p{1.2cm}!{\vrule width 0.8pt}p{0.8cm}!{\vrule width 0.8pt}}
\Xhline{3\arrayrulewidth}
\textbf{Mechanism} & \textbf{Description} & \textbf{Core Operation} & \textbf{Modality} & \textbf{Ref.} \\
\Xhline{3\arrayrulewidth}
Gaussian Noise & Adds zero-mean Gaussian noise independently to each frame. & $x_t^{'} = x_t + \mathcal{N}(0, \sigma^2)$ & Eye, Body & \cite{david-john2021Privacypreservingc, wilson2024PrivacyPreserving} \\
\Xhline{0.8pt}
Temporal Downsampling & Reduces sampling rate by keeping every $K$-th frame and duplicating information from previous frames & $x_t^{'} = x_{t-K}$ if $t \mod K \neq 0$. & Eye, Body & \cite{wilson2024PrivacyPreserving} \\
\Xhline{0.8pt}
Spatial Downsampling & Maps high-resolution spatial information to discrete bins or grid cells & $x_t^{'} = \text{round}(x_t / \delta) \cdot \delta$. & Eye, Body & \cite{david-john2021Privacypreservingc, wilson2024PrivacyPreserving} \\
\Xhline{0.8pt}
Smoothing & Averages current and $B$ previous samples using linearly decreasing weights. & $x_t^{'} = \frac{1}{Z} \sum_{i=1}^{B} w_i x_{t-i}$ where $w_i \propto i$ & Eye, Body & \cite{wilson2024PrivacyPreserving} \\
\Xhline{0.8pt}
Body-LDP & Adds bounded Laplace noise to height, wingspan, and joint positions.   & $x_t^{'} = x_t + \text{Lap}(b)$ with bounded domain & Body & \cite{nair2023Going, kasiviswanathan_what_2011} \\
\Xhline{0.8pt}
DMM & Uses a neural network to perturb motion while preserving realism, trained adversarially against an identity discriminator & $x^{'} = f_{\text{DMM}}(x)$ via adversarial training. & Body & \cite{nair2023deep} \\
\Xhline{0.8pt}
Composite Mechanism & Applies two privacy mechanisms in sequence to  {harness} their complementary strengths & $x^{'} = M_2(M_1(x))$. & Body & Ours \\
\Xhline{3\arrayrulewidth}
\end{tabular}
\label{tab:privacy_mechanisms}
\end{table*}
}
\begin{figure*}[]
    \centering
    \includegraphics[width=0.85\linewidth]{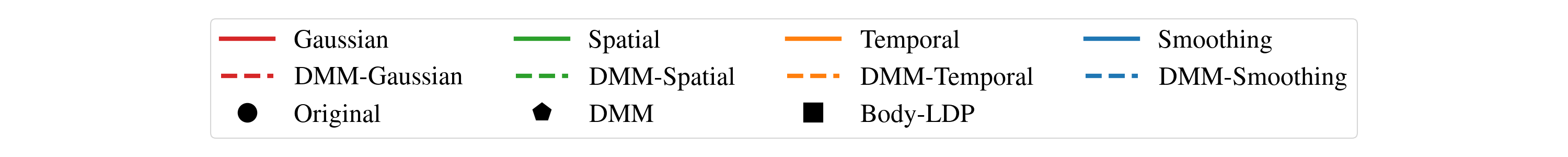}
    
    \begin{subfigure}[h]{0.33\linewidth}
        \includegraphics[width=\linewidth]{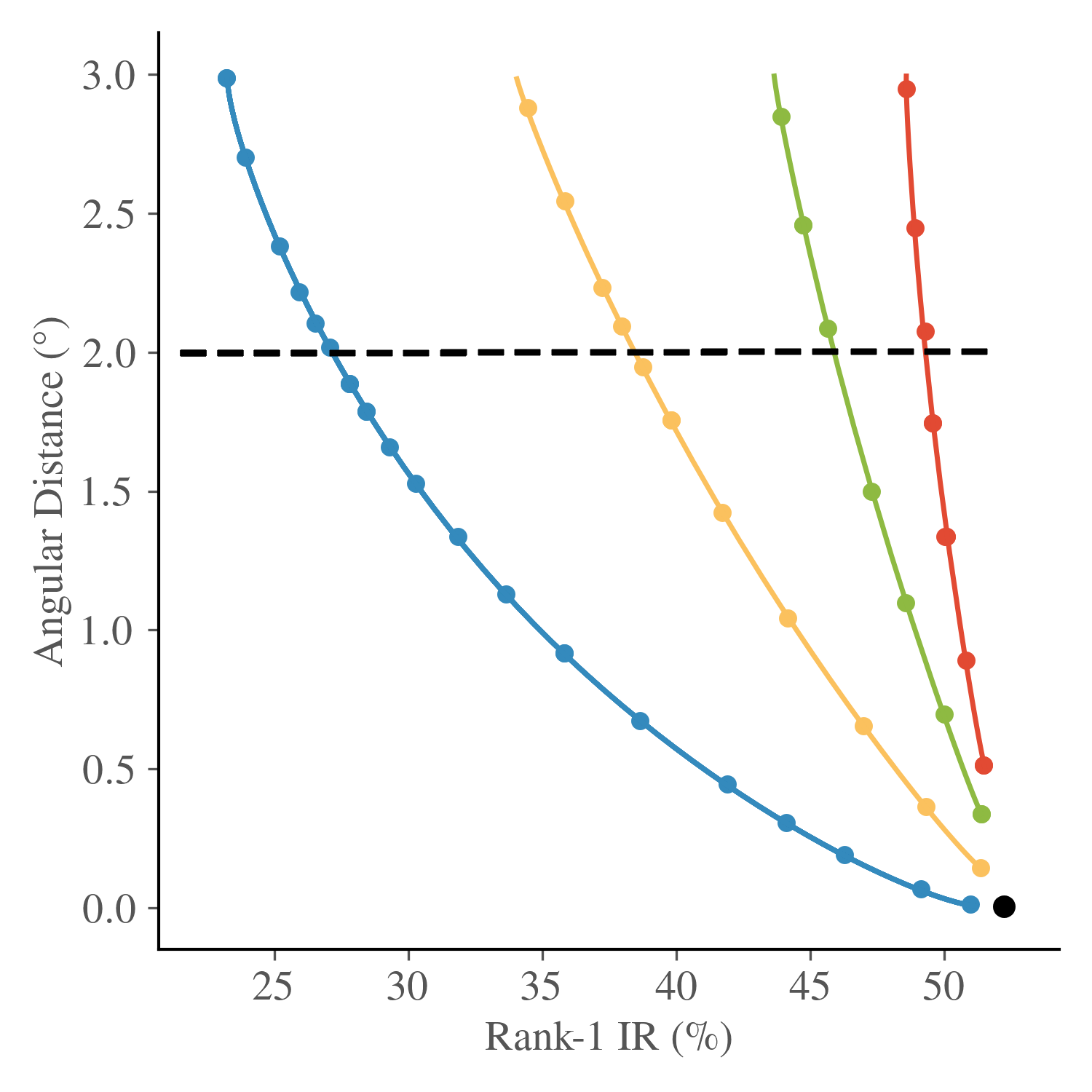}
        % \caption{Privacy-utility trade-off for eye movement mechanisms. Smoothing achieves the best balance, reducing identification accuracy while introducing minimal angular distortion.}
        \caption{Eye  {Telemetry}}
        \label{fig:eye_privacy_utility}
    \end{subfigure}
    \begin{subfigure}[h]{0.33\linewidth}
        \includegraphics[width=\linewidth]{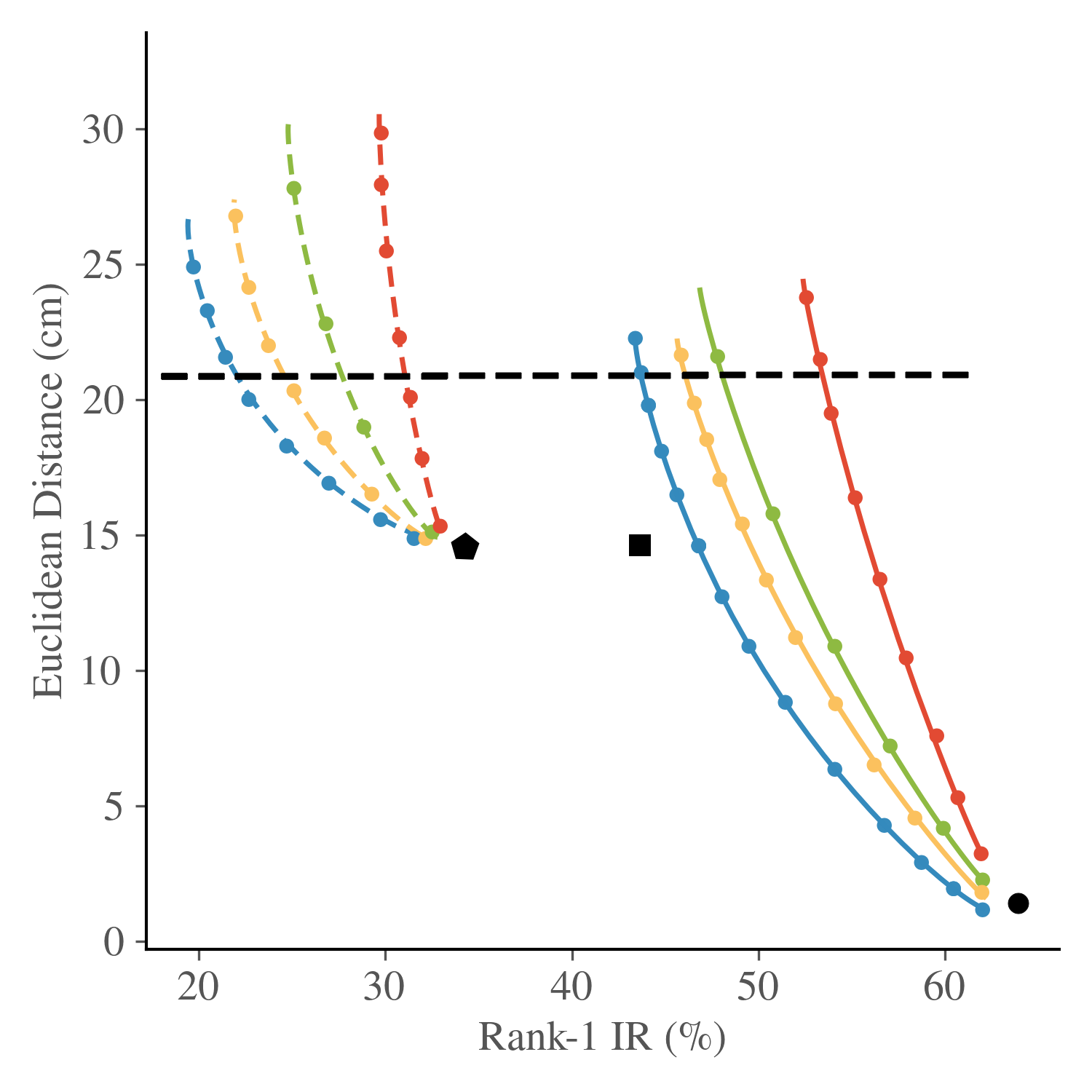}
        % \caption{Privacy-utility trade-off for body movement mechanisms. \textit{Smoothing+DMM} achieves the best results, offering strong anonymization while preserving motion realism. Only the top four composite mechanisms are shown.}
        \caption{Body  {Telemetry}}
        \label{fig:body_privacy_utility}
    \end{subfigure}
    \caption{Privacy-utility trade-off comparison for eye (left) and body (right)  {privacy} mechanisms in Causal XR applications. Each plot presents the impact of different privacy methods on re-identification performance, measured by Rank-1 identification rate (IR) on the x-axis, and utility degradation, measured by angular distance (left) or Euclidean distance (right) on the y-axis. \textbf{Points to the left indicate stronger privacy (lower re-identification rates), points toward the bottom indicate better utility preservation (minimal distance from original), and the bottom-left region represents optimal privacy-utility balance}. Solid lines represent standalone mechanisms, while dashed lines represent DMM-based composite mechanisms. The original data points are indicated by circles. Dashed black lines represent literature thresholds shown to preserve interaction fidelity~\cite{schuetz2020psychophysics, wentzel2020improving}. }
    \label{fig:privacy_utility_combined}
\end{figure*}

We now examine re-identification risks in casual XR applications. These include experiences such as Job Simulator\cite{jobsimulator} and VRChat\cite{vrchat}, which often involve tasks like object manipulation, object selection and virtual browsing. Such interactions prioritize spatial accuracy over temporal responsiveness. In these scenarios, small delays introduced by privacy mechanisms may be acceptable, as long as interaction precision is preserved. The unique characteristics of these applications lead us to pose our first research question:

 \textit{RQ1: Among eye and body privacy mechanisms, which provide the best trade-off between re-identification risk and interaction fidelity in casual XR applications?}

\subsection{Utility Metric}

As mentioned earlier, casual XR applications emphasize spatial accuracy over temporal responsiveness. To evaluate the impact of anonymization mechanisms in this context, we use modality-specific utility metrics that reflect spatial and directional accuracy during tasks like virtual browsing, object manipulation and selection.

 \textbf{Eye  {Telemetry} – Angular Distance.}
Gaze direction plays a critical role in many XR interfaces for object selection and interaction. To quantify the effect of anonymization on gaze precision, we use angular distance, which measures the angle between the original and  {privatized} gaze vectors. This metric captures directional deviation without being influenced by absolute position, aligning well with spatial gaze-based interactions. Smaller angular distances indicate minimal distortion and higher interaction fidelity.   {We compute angular distance for each frame and report the average across all frames: }

\begin{equation}
\theta = \frac{1}{T} \sum_{t=1}^{T} \arccos{\left( \frac{\vec{v}_1^{(t)} \cdot \vec{v}_2^{(t)}}{|\vec{v}_1^{(t)}| |\vec{v}_2^{(t)}|} \right)}
\end{equation}

 where $\vec{v}_1^{(t)}$ and $\vec{v}_2^{(t)}$ denote the original and anonymized gaze direction vectors at time frame $t$, respectively. Prior literature reports that angular distance below $2^\circ$ preserves interaction fidelity in spatially precise gaze tasks~\cite{schuetz2020psychophysics}

 \textbf{Body  {Telemetry} – Euclidean Distance.}
Body movement data is essential for interactions such as reaching, pointing, and manipulating objects. To evaluate spatial fidelity after anonymization, we compute the Euclidean distance between the original and anonymized positions of key body joints. This metric reflects physical displacement and is well-suited to assess whether natural interaction is preserved.  {We calculate the Euclidean distance independently for the headset, left controller, and right controller at each time frame, and report the average distance across all frames:}

\ifshowfeedback
    \begin{equation}
    \stkout{d} = \frac{\stkout{1}}{\stkout{T}} \sum_{t=1}^{T} \frac{
    \stkout{\|\mathbf{h}_1^{(t)} - \mathbf{h}_2^{(t)}\| +
    \|\mathbf{l}_1^{(t)} - \mathbf{l}_2^{(t)}\| +
    \|\mathbf{r}_1^{(t)} - \mathbf{r}_2^{(t)}\|}
    }{\stkout{3}}
    \end{equation}
\textcolor{darkgreen}{
\begin{equation}
    d = \frac{1}{T} \sum_{t=1}^{T} \frac{
    \|\mathbf{h}_1^{(t)} - \mathbf{h}_2^{(t)}\|_2 +
    \|\mathbf{l}_1^{(t)} - \mathbf{l}_2^{(t)}\|_2 +
    \|\mathbf{r}_1^{(t)} - \mathbf{r}_2^{(t)}\|_2
    }{3}
    \end{equation}
}
\else
\begin{equation}
    d = \frac{1}{T} \sum_{t=1}^{T} \frac{
    \|\mathbf{h}_1^{(t)} - \mathbf{h}_2^{(t)}\|_2 +
    \|\mathbf{l}_1^{(t)} - \mathbf{l}_2^{(t)}\|_2 +
    \|\mathbf{r}_1^{(t)} - \mathbf{r}_2^{(t)}\|_2
    }{3}
    \end{equation}

\fi

\ethanwilsonroundone{This isn't the formula for Euclidean distance!  This is MAE}
\aibragimovroundone{Thanks for noticing this - completely slipped out of mind that not specifying a number next to ||, by default, means L1 norm. I changed the number to indicate this is L2 norm (aka euclidian distance). }

 \noindent \textcolor{black}{where $\mathbf{h}_1^{(t)}$, $\mathbf{l}_1^{(t)}$, $\mathbf{r}_1^{(t)}$ and $\mathbf{h}_2^{(t)}$, $\mathbf{l}_2^{(t)}$, $\mathbf{r}_2^{(t)}$ represent the original and anonymized 3D positions of the headset, left controller, and right controller at time frame $t$, respectively. Prior research reports that Euclidean distance threshold of 21 cm  preserves spatial task fidelity in XR environments~\cite{wentzel2020improving}. }

To defend against re-identification in XR, we evaluate a set of unimodal  {mechanisms, including standalone} and composite privacy mechanisms, summarized in Table~\ref{tab:privacy_mechanisms}. These include  {} approaches such as Gaussian noise addition, temporal and spatial downsampling, and smoothing, as well as  {newer} defenses like Body-LDP and Deep Motion Masking (DMM). Each mechanism targets either eye or body signals independently and is drawn from prior work focused on real-time privacy preservation in immersive environments. In addition, we propose a \textit{composite mechanism} that sequentially applies two unimodal body privacy methods to enhance protection while preserving usability.

\ethanwilsonroundone{I don't think algorithm pseudocode is needed to explain a simple sequence.}

\ejainroundone{It's simple but I would keep it. I found it useful. Else replace with block diagram.}

\aibragimovroundone{I agree with Dr. Jain here - it might be useful for readers even though its very simple. }

% \begin{algorithm}[H]
% \caption{Composite Privacy Mechanism}
% \begin{algorithmic}[1]
% \Require $D$: Original body movement data record
% \Ensure $D_{composite}$: Composite privatized record

% \State $D_1 \gets \texttt{Mechanism}_1(D)$
% \State $D_2 \gets \texttt{Mechanism}_2(D_1)$
% \State \Return $D_2$
% \end{algorithmic}
% \label{alg:composite}
% \end{algorithm}

% In our evaluation, we test various combinations of body privacy mechanisms by varying $\texttt{Mechanism}_1$ and $\texttt{Mechanism}_2$ across available options. We then analyze their composite impact on privacy (via re-identification accuracy) and utility (via Euclidean distance), identifying the pairs that achieve the best trade-offs.

\subsection{Dataset}

For our analysis, we use the OpenNEEDS dataset~\cite{emery2021openneeds}, which is designed to represent casual XR usage. The dataset includes data from 44 participants engaged in low-intensity, slow-paced tasks such as reading, drawing, and object manipulation. These tasks emphasize spatial precision over reaction time or movement speed. 

The dataset includes synchronized eye and body  {telemetry} signals collected via XR headsets and hand-held controllers from 44 participants. To ensure data quality and consistency, we excluded participants with fewer than two sessions or less than 30 seconds of valid recordings. Our final analysis includes 38 participants, each contributing at least two valid sessions.

\subsection{Evaluation Methodology and Identification Protocol}

 \textbf{Experimental Setup.}
We assess the effectiveness of the proposed privacy mechanisms through re-identification experiments using the EKYT~\cite{lohr2022Eyeb, aziz2024exploiting} model under two modalities: eye movements and body movements. This model is particularly suitable for our evaluation, as prior work by Aziz et al.~\cite{aziz2024exploiting} demonstrated that it can be easily adapted to eye-only, body-only, and multimodal inputs. Additionally, it has been previously used for privacy evaluations in XR settings, making it well-aligned with the goals of our study~\cite{aziz2024exploiting}. A 4-fold cross-validation protocol is employed, where each model is trained on 75\% of the participants and evaluated on the remaining 25\%. Test folds are non-overlapping to ensure participant-level separation between training and evaluation.

To quantify re-identification performance, we adopt a two-session matching protocol. For each test participant, the first session is used to generate gallery embeddings, while the second session provides probe embeddings. Cosine similarity is computed between each probe and all gallery embeddings, and performance is reported as Rank-1 Identification Rate (Rank-1 IR), the percentage of correct top-1 matches. Privacy mechanisms are applied to both probe and gallery data, simulating a strong adversary with access to privatized data or the ability to apply equivalent transformations to incoming samples.

 \textbf{Privacy–Utility Trade-Off.}
As privacy mechanisms can introduce distortions that affect user interaction, we systematically evaluate the trade-off between privacy preservation and utility retention. For each mechanism, we sweep across a range of parameter values and identify configurations that offer the best balance, maximizing privacy while minimizing degradation in utility metrics. This analysis is conducted independently for each modality using their respective utility metrics.

\subsection{Results}

\ethanwilsonroundone{In the abstract / introduction it's stated multiple times that the mechanisms don't violate real-time usability constraints.  But it's not clear yet at this point in the paper what those are.  What angular or Euclidean distance is acceptable, and why?}

\aibragimovroundone{This a good point. I put a notion of this in introduction}

Figure~\ref{fig:privacy_utility_combined} illustrates the privacy–utility trade-offs for various anonymization mechanisms applied to eye and body movement data in casual XR applications. The left subplot corresponds to eye  {telemetry}, and the right subplot to body  {telemetry}. Each curve represents a mechanism evaluated across a range of parameter values. Re-identification risk is measured using Rank-1 IR on the x-axis, while utility degradation is quantified using angular distance (for gaze  {telemetry}) and Euclidean distance (for body  {telemetry}) on the y-axis. Solid lines represent standalone privacy mechanisms, while dashed lines indicate DMM-based composites where the DMM mechanism is applied together with another privacy method such as Gaussian noise, temporal downsampling, or spatial downsampling. Dashed black lines represent literature thresholds shown to preserve interaction fidelity~\cite{schuetz2020psychophysics, wentzel2020improving}

 \textbf{Eye  {Privacy} Mechanisms.}  
Among all evaluated methods, Smoothing consistently achieves the most favorable utility–privacy trade-off for eye  {telemetry}, reaching the optimal region in the trade-off space. In comparison, Gaussian noise yields moderate privacy improvements but introduces substantial angular distortion as the noise level increases. Temporal and Spatial downsampling follow intermediate patterns, though both result in greater  {} deviation than smoothing.

 \textbf{Body Privacy Mechanisms.}  
For body  telemetry, composite mechanisms consistently outperform standalone methods. The  {composition} of DMM-Smoothing achieves the most favorable trade-off, significantly reducing Rank-1 IR while preserving spatial fidelity. DMM-Temporal and DMM-Gaussian also demonstrate strong performance, offering a balanced reduction in identifiability with minimal degradation in movement quality. 

In contrast, standalone mechanisms such as Gaussian noise, Temporal downsampling, and Spatial downsampling provide either limited privacy benefits or induce considerable motion distortion. The Body-LDP baseline performs moderately but is consistently outperformed by  composites of DMM and others.

 Considering the results discussed in this section, we arrive at the following answer to RQ1: 
 
 \textit{In casual XR applications, we find that Smoothing for eye  telemetry and DMM-Smoothing for body  telemetry achieve the most favorable trade-off between re-identification risk and interaction fidelity.}

\section{Evaluation of Anonymization in Competetive XR Applications}

We now examine re-identification risks in competitive XR applications. These include high-intensity experiences that demand low latency and fast response times. Even minor delays can significantly degrade user performance. Typical tasks involve rhythm-based gesture execution, timed object interception, and rapid target acquisition under strict timing constraints. These interaction patterns are common in applications, such as Beat Saber\cite{beatsaber} and Synth Riders\cite{synthriders}. The unique properties of these applications lead us to pose our second research question:

    \textit{RQ2: In competitive XR applications, which eye and body privacy mechanisms most effectively balance re-identification protection with the competitive interaction demands required for high-performance user experiences?}

\subsection{Utility Metric}

In competitive XR environments, utility is closely linked to  {low-latency interactions, where fast response times are critical for task success}. To evaluate utility under the competitive conditions, we use modality-specific metrics that capture the impact of privacy mechanisms on time-sensitive task performance.

 \textbf{Eye  {Privacy} – Area of Interest Accuracy.}
To assess the utility of eye  {telemetry}, we adopt the Area of Interest (AOI) accuracy metric used by Wilson et al.~\cite{wilson2024PrivacyPreserving}, who extensively evaluated the impact of various privacy mechanisms on user performance in competitive XR tasks. Their analysis identified optimal anonymization techniques and parameter settings that preserve interaction fidelity while reducing re-identification risk. Accordingly, we use the same eye  {privacy} mechanism and parameters in our evaluation, and do not re-explore the privacy–utility trade-off for this modality. Our focus instead is on body and multimodal mechanisms.

 \textbf{Body  {Privacy} – Beat Saber Score.}
In competitive XR applications, player performance is often quantified using in-game scores that reflect task completion accuracy and timing. These scores are influenced by motion intensity {and responsiveness}. To measure the impact of privacy mechanisms on gameplay performance, we use \textit{SimSaber}~\cite{cull2024simsaber}, a Beat Saber replay analysis tool that simulates gameplay and computes scores based on  {telemetry} recordings. This allows for a direct comparison between anonymized and original motion data. The score difference is defined as:

\begin{equation}
S = T(\mathbf{v}_1) - T(\mathbf{v}_2)
\end{equation}

 where $T(\cdot)$ denotes the scoring function, $\mathbf{v}_1$ is the original motion trace, and $\mathbf{v}_2$ is the anonymized version. Smaller values of $S$ indicate better preservation of gameplay performance. The prior literature shows that a threshold of 80\% of the original Beat Saber score reflects acceptable performance degradation for high-precision gameplay tasks~\cite{nair2023deep}.

\subsection{Dataset}

\textbf{Chimera.} Competitive XR applications frequently store detailed telemetry online for extended periods. Leaderboard platforms such as BeatLeader, for example, maintain motion traces from competitive gameplay that can span months or even years~\cite{nair2023Uniquea}. Over these timescales, both eye and body movement patterns can change significantly due to physical growth, skill development, fatigue, and evolving hardware~\cite{griffith2021GazeBasea, nair2023Uniquea}. This drift is especially pronounced in children, who are a large demographic for XR gaming, and whose motor control, posture, and gaze behaviors may differ dramatically from one year to the next~\cite{malone2025control, mareschal2016you, griffith2021GazeBasea}.

From a privacy perspective, such longitudinal changes create a \textit{toughest-case privatization scenario}. In this scenario, the fine-grained, moment-to-moment coordination between modalities has eroded, yet subtle, modality-specific biometric traits remain. A robust privacy mechanism must still prevent re-identification under these conditions, where defending against linkage is most challenging despite the attacker’s reduced access to synchronous cues.

To model this scenario at scale, we introduce the \textit{Chimera dataset}. This dataset is built using an established chimeric pairing approach from biometric research~\cite{ross2004multimodal, connaughton2013fusion, poh2006chimeric}. Eye telemetry comes from GazeBaseVR~\cite{lohr2023GazeBaseVR}, which records high-resolution gaze data during structured visual tasks, and body telemetry from BeatLeader~\cite{nair2023Uniquea}, which captures competitive Beat Saber motion traces. Both datasets involve high engagement and movement intensity, ensuring plausibility when fused. By pairing participants across datasets (Algorithm~\ref{alg:chimera}), we simulate the \textit{toughest-case privatization scenario} in which a user’s gaze and body movement patterns have changed so substantially over time that little remains of their original fine-grained coordination. The resulting 407 chimeric identities therefore reflect the most privacy-challenging form of long-term archival data, where persistent biometric traits exist but session-to-session behavioral overlap is minimal.

\textbf{Wilson et al.} While Chimera targets the long-term drift regime as the toughest-case privatization scenario, it is equally important to study the opposite extreme. This extreme involves short-term, high-correlation multimodal data where gaze and body cues are tightly synchronized. For this reason, we evaluate re-identification risks on the real-world dataset of Wilson et al. (Section~6), which contains naturally synchronized eye and body telemetry from 26 participants performing a competitive XR cooking task. Together, Chimera and Wilson span the full multimodal correlation spectrum, allowing us to test privacy mechanisms against both worst-case longitudinal drift and immediate-session coordination.

\begin{algorithm}[t]
\footnotesize
\caption{Creating a Multimodal Chimera Dataset}
\begin{algorithmic}[1]
    \Require Lists of identities from GazeBaseVR $G$, BeatLeader $B$
    \Ensure List of multimodal chimera identity pairs $C$
    \State $C \gets [\,]$ 
    \While{$G \neq [\,]$ \textbf{and} $B \neq [\,]$}
        \State $i \gets \text{random integer in } [0, \text{len}(G) - 1]$
        \State $j \gets \text{random integer in } [0, \text{len}(B) - 1]$
        \State $g \gets G.\text{pop}(i)$ 
        \State $b \gets B.\text{pop}(j)$ 
        \State $C.\text{append}((g, b))$ 
    \EndWhile
    \State \Return $C$
\end{algorithmic}
\label{alg:chimera}
\end{algorithm}

\subsection{Evaluation Methodology}

\textbf{Experimental Setup \& Identification Protocol.}
We evaluate the effectiveness of the proposed privacy mechanisms through re-identification experiments using the EKYT model across body  {telemetry}. A 4-fold cross-validation protocol is used, with the train set comprising 361 participants for training, and the test set containing 46 participants for testing.  We adopt a two-session matching protocol to measure identification performance. For each test participant, gallery embeddings are generated from the first session and probe embeddings from the second. Cosine similarity is computed between probe-gallery pairs, and identification accuracy is reported as Rank-1 IR, expressed as a percentage. Privacy mechanisms are applied to both probe and gallery samples, simulating a strong adversary with access to privatized data or the ability to apply identical transformations at inference time.

 \textbf{Privacy Mechanisms Evaluated.}
In this analysis, we evaluate privacy mechanisms applied to body  {telemetry}, including Gaussian noise, temporal downsampling, spatial downsampling, smoothing, Deep Motion Masking (DMM), Body-LDP, and composite methods that  {compose} DMM with other techniques. We assess the privacy-utility trade-offs of these approaches to understand their impact on re-identification risk and motion fidelity for competitive XR applications.

 \textbf{Privacy–Utility Trade-Off Analysis.}
Privacy mechanisms often introduce distortions that can degrade system performance and user experience. To evaluate the balance between privacy protection and utility preservation, we systematically vary mechanism parameters and measure both identification accuracy and task-relevant utility degradation. For each mechanism, we identify configurations that offer strong anonymization while maintaining acceptable utility levels in unimodal and multimodal scenarios.

% \textbf{Threshold-Based Parameter Selection.}  
% To ensure fairness and maintain real-world usability, we select mechanism configurations based on distortion thresholds grounded in prior VR and HCI literature. These thresholds represent the maximum allowable degradation before user performance is expected to decline:

% \begin{itemize}
%     \item \textbf{Body movements:} Beat Saber gameplay performance drops of up to \textbf{50\%} are considered tolerable for preserving task completion and engagement.
% \end{itemize}

% All privacy mechanisms are tuned to operate at or just below these thresholds. This standardization enables consistent, usability-aware comparisons across all evaluated modalities and privacy configurations.

\begin{figure}[]
    \centering
    \includegraphics[width=0.7\linewidth]{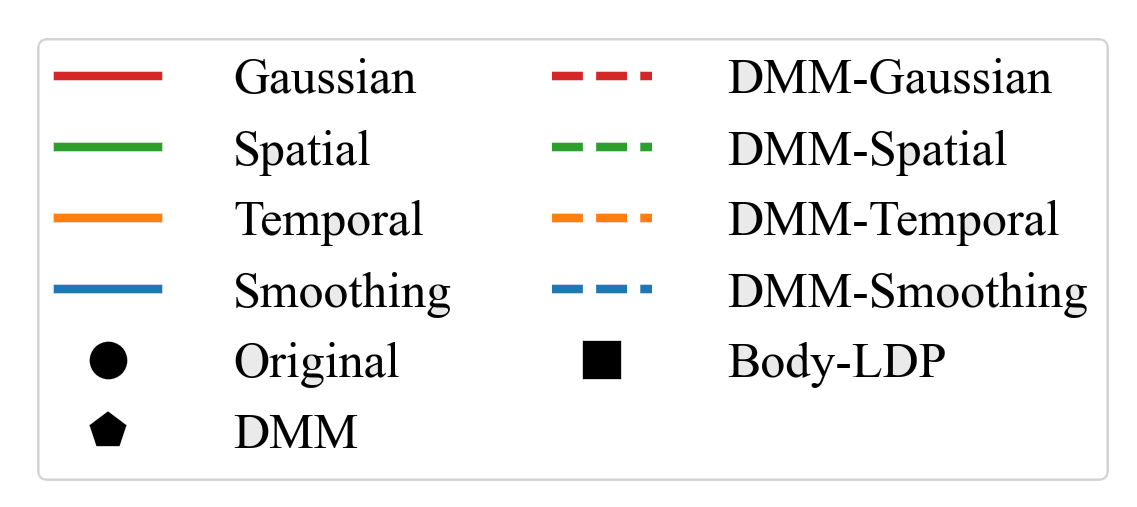}

    % \begin{subfigure}[t]{0.49\linewidth}
    %     \includegraphics[width=\linewidth]{figures/euclidean_vs_accuracy_eyes (1).png}
    %     % \caption{Privacy-utility trade-off for eye movement mechanisms. Smoothing achieves the best balance, reducing identification accuracy while introducing minimal angular distortion.}
    %     \label{fig:eye_privacy_utility}
    % \end{subfigure}
    \includegraphics[width=0.7\linewidth]{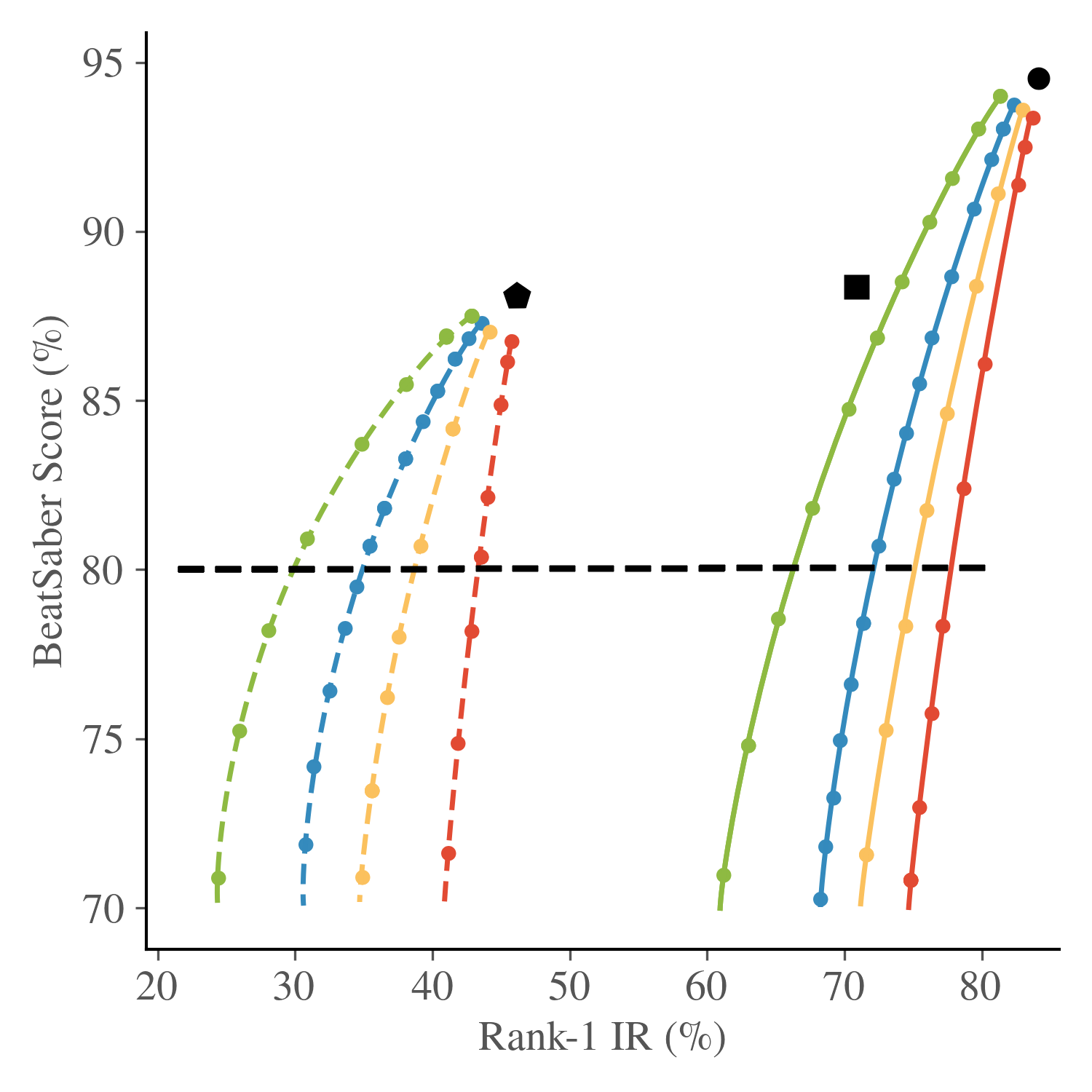}
        % \caption{Privacy-utility trade-off for body movement mechanisms. \textit{Smoothing+DMM} achieves the best results, offering strong anonymization while preserving motion realism. Only the top four composite mechanisms are shown.}
    \caption{Comparison of privacy-utility trade-offs for body  telemetry mechanisms. The plot illustrates how different privacy methods affect Rank-1 identification rate (IR) and the corresponding utility, measured by BeatSaber Score. \textbf{Points to the left indicate stronger privacy (lower Rank-1 IR), points toward the top indicate better utility preservation (maximum preservation of BeatSaber score), and the bottom-left region represents optimal privacy-utility balance}. Solid lines represent standalone mechanisms, while dash lines represent DMM-based composite mechanisms. The original data points are indicated by circles. Dashed black line represents literature thresholds shown to preserve interaction fidelity~\cite{nair2023deep}.
 }
\label{fig:results_competetive}
\end{figure}

\begin{table*}[!htbp]
\centering
\tiny
\caption{Rank-1 IR (\%) on the \textbf{OpenNEEDS dataset (casual XR application)} for various  {pairs} of eye and body  {} privacy mechanisms. Each cell reports Gaze-based, Body-based, and Multimodal identification rates in the format: Gaze IR / Body IR / Multimodal IR. Lower values indicate stronger privacy protection and reduced re-identification risk. Bolded values indicate results that provide the strongest privacy protection within each category and underlined values indicate best overall.
}
\renewcommand{\arraystretch}{0.95}
\resizebox{\textwidth}{!}{
\begin{tabular}{!{\vrule width 0.8pt}l!{\vrule width 0.8pt}c:c:c:c:c!{\vrule width 0.8pt}}
\multicolumn{1}{c}{} & \multicolumn{5}{c}{\textbf{\underline{OpenNEEDS (Casual XR Application)}}} \\
\Xhline{3\arrayrulewidth}
\diagbox[width=\dimexpr \textwidth/10+2\tabcolsep\relax, height=0.33cm]{\textbf{\textit{Body Mech.}}}{\textbf{\textit{Eye Mech.}}}
& \textit{None} & \textit{Gaussian} & \textit{Spatial} & \textit{Temporal} & \textit{Smoothing} \\
\Xhline{3\arrayrulewidth}
\textit{None} & 52.6 / 72.4 / 80.3 & 48.7 / 72.4 / 78.9 & 46.7 / 72.4 / 76.3 & 36.2 / 72.4 / 73.0 & \textbf{27.6 / 72.4 / 72.9} \\
\textit{Gaussian} & 52.6 / 59.2 / 63.2 & 48.7 / 59.2 / 63.2 & 46.7 / 59.2 / 63.2 & 36.2 / 59.2 / 62.5 & \textbf{27.6 / 59.2 / 61.2} \\

\textit{Spatial} & 52.6 / 53.3 / 58.6 & 48.7 / 53.3 / 58.6 & 46.7 / 53.3 / 56.6 & 36.2 / 53.3 / 50.0 & \textbf{27.6 / 53.3 / 47.4} \\
\textit{Temporal} & 52.6 / 50.0 / 55.3 & 48.7 / 50.0 / 52.6 & 46.7 / 50.0 / 52.0 & 36.2 / 50.0 / 51.3 & \textbf{27.6 / 50.0 / 50.0} \\
\textit{Smoothing} & 52.6 / 48.7 / 54.6 & 48.7 / 48.7 / 50.0 & 46.7 / 48.7 / 49.3 & 36.2 / 48.7 / 46.7 & \textbf{27.6 / 48.7 / 46.1} \\
\textit{Body-LDP} & 52.6 / 48.6 / 54.6 & 48.7 / 48.6 / 49.2 & 46.7 / 48.6 / 49.2 & 36.2 / 48.6 / 46.1 & \textbf{27.6 / 48.6 / 46.1} \\

\textit{DMM} & 52.6 / 37.5 / 41.4 & 48.7 / 37.5 / 40.1 & 46.7 / 37.5 / 40.1 & 36.2 / 37.5 / 39.5 & \textbf{27.6 / 37.5 / 39.5} \\
\textit{DMM-Gaussian} & 52.6 / 33.6 / 39.5 & 48.7 / 33.6 / 38.8 & 46.7 / 33.6 / 38.2 & 36.2 / 33.6 / 37.5 & \textbf{27.6 / 33.6 / 36.2} \\
\textit{DMM-Spatial} & 52.6 / 27.6 / 38.8 & 48.7 / 27.6 / 36.2 & 46.7 / 27.6 / 35.5 & 36.2 / 27.6 / 34.2 & \textbf{27.6 / 27.6 / 33.6} \\
\textit{DMM-Temporal} & 52.6 / 24.3 / 36.2 & 48.7 / 24.3 / 33.6 & 46.7 / 24.3 / 33.6 & 36.2 / 24.3 / 31.6 & \textbf{27.6 / 24.3 / 27.6} \\
\textit{DMM-Smoothing} & \textbf{52.6 / 21.1 / 33.6} & \textbf{48.7 / 21.1 / 31.6} & \textbf{46.7 / 21.1 / 31.6} & \textbf{36.2 / 21.1 / 29.6} & \textbf{\ul{27.6 / 21.1 / 26.3}} \\
\bottomrule
\end{tabular}
}
\label{tab:gaze_body_multireid}
\end{table*}

\subsection{Results}

Figure~\ref{fig:results_competetive} presents the privacy-utility trade-off for various anonymization mechanisms applied to body  {telemetry} in competitive XR scenarios. Re-identification risk is measured using Rank-1 IR, while utility is measured using Beat Saber score, which reflects user performance in a fast-paced, task-sensitive environment.

 \textbf{Body  {Privacy} Mechanisms.}  
When evaluating task performance with body telemetry, composite mechanisms again yield the most favorable trade-offs. DMM-Spatial stands out by achieving a strong balance between re-identification resistance and gameplay utility, maintaining high Beat Saber scores while reducing Rank-1 IR. DMM-Temporal and DMM-Smoothing also perform well, offering privacy gains with minimal impact on performance. In contrast, standalone mechanisms show weaker trade-offs, either preserving utility at the cost of higher identifiability or degrading task performance more severely. The Body-LDP baseline offers a moderate compromise but is consistently outperformed by DMM-based composites.

% \begin{figure}[h]
%     \centering
%     \includegraphics[width=0.9\linewidth]{figures/body_tradeoff_competetive.png}
%     \caption{Caption}
%     \label{fig:body_tradeoff_competetive}
% \end{figure}
\setul{0.5ex}{0.3ex} % First arg = distance from baseline, second = line thickness

\begin{table*}[!htbp]
\tiny
\caption{Rank-1 IR (\%) on the \textbf{Chimera and Wilson et al. datasets (competitive XR application)} for various  {pairs} of eye and body  {} privacy mechanisms. Each cell reports Gaze-based, Body-based, and Multimodal identification rates in the format: Gaze IR / Body IR / Multimodal IR. Lower values indicate stronger privacy protection and reduced re-identification risk. Bolded values indicate results that provide the strongest privacy protection within each category and underlined values indicate best overall.
}
\renewcommand{\arraystretch}{0.95}
\resizebox{\textwidth}{!}{
\begin{tabular}{!{\vrule width 0.8pt}l!{\vrule width 0.8pt}c:c:c:c:c!{\vrule width 0.8pt}}
\multicolumn{1}{c}{} & \multicolumn{5}{c}{\textbf{\underline{Chimera (Competitive XR Application)}}} \\
\Xhline{3\arrayrulewidth}
\diagbox[width=\dimexpr \textwidth/10+2\tabcolsep\relax, height=0.33cm]{\textbf{\textit{Body Mech.}}}{\textbf{\textit{Eye Mech.}}}
& \textit{None} & \textit{Gaussian} & \textit{Spatial} & \textit{Temporal} & \textit{Smoothing} \\
\Xhline{3\arrayrulewidth}
\textit{None} & 50.0 / 80.4 / 84.8 & 34.8 / 80.4 / 82.6 & 26.1 / 80.4 / 73.9 & 45.7 / 80.4 / 82.6 & \textbf{23.9 / 80.4 / 80.4} \\
\textit{Gaussian} & 50.0 / 76.1 / 80.4 & 34.8 / 76.1 / 78.3 & 26.1 / 76.1 / 76.1 & 45.7 / 76.1 / 80.4 & \textbf{23.9 / 76.1 / 76.1} \\
\textit{Spatial} & 50.0 / 63.0 / 71.7 & 34.8 / 63.0 / 71.7 & 26.1 / 63.0 / 58.7 & 45.7 / 63.0 / 69.6 & \textbf{23.9 / 63.0 / 65.2} \\
\textit{Temporal} & 50.0 / 71.7 / 78.3 & 34.8 / 71.7 / 76.1 & 26.1 / 71.7 / 67.4 & 45.7 / 71.7 / 73.9 & \textbf{23.9 / 71.7 / 71.7} \\
\textit{Smoothing} & 50.0 / 69.6 / 76.1 & 34.8 / 69.6 / 73.9 & 26.1 / 69.6 / 63.0 & 45.7 / 69.6 / 71.7 & \textbf{23.9 / 69.6 / 69.6} \\
\textit{Body-LDP} & 50.0 / 65.2 / 71.7 & 34.8 / 65.2 / 69.6 & 26.1 / 65.2 / 58.7 & 45.7 / 65.2 / 67.4 & \textbf{23.9 / 65.2 / 65.2} \\
\textit{DMM} & 50.0 / 41.3 / 41.3 & 34.8 / 41.3 / 41.3 & 26.1 / 41.3 / 39.1 & 45.7 / 41.3 / 41.3 & \textbf{23.9 / 41.3 / 41.3} \\
\textit{DMM-Gaussian} & 50.0 / 39.1 / 39.1 & 34.8 / 39.1 / 39.1 & 26.1 / 39.1 / 36.9 & 45.7 / 39.1 / 39.1 & \textbf{23.9 / 39.1 / 41.3} \\
\textit{DMM-Spatial} & \textbf{50.0 / 26.1 / 30.4} & \textbf{34.8 / 26.1 / 28.3} & \textbf{26.1 / 26.1 / 28.3} & \textbf{45.7 / 26.1 / 30.4} & \textbf{\ul{23.9 / 26.1 / 26.1}} \\
\textit{DMM-Temporal} & 50.0 / 32.6 / 36.9 & 34.8 / 32.6 / 36.9 & 26.1 / 32.6 / 32.6 & 45.7 / 32.6 / 36.9 & \textbf{23.9 / 32.6 / 39.1} \\
\textit{DMM-Smoothing} & 50.0 / 30.4 / 30.4 & 34.8 / 30.4 / 30.4 & 26.1 / 30.4 / 30.4 & 45.7 / 30.4 / 32.6 & \textbf{23.9 / 30.4 / 30.4} \\
\bottomrule
\end{tabular}
}
\renewcommand{\arraystretch}{0.95}
\resizebox{\textwidth}{!}{
\begin{tabular}{!{\vrule width 0.8pt}l!{\vrule width 0.8pt}c:c:c:c:c!{\vrule width 0.8pt}}
\multicolumn{1}{c}{} & \multicolumn{5}{c}{\textbf{\underline{Wilson et al. (Competitive XR Application)}}} \\
\Xhline{3\arrayrulewidth}
\diagbox[width=\dimexpr \textwidth/10+2\tabcolsep\relax, height=0.35cm]{\textbf{\textit{Body Mech.}}}{\textbf{\textit{Eye Mech.}}}
& \textit{None} & \textit{Gaussian} & \textit{Spatial} & \textit{Temporal} & \textit{Smoothing} \\
\Xhline{3\arrayrulewidth}
\textit{None} & 61.5 / 88.5 / 92.3 & 42.3 / 88.5 / 92.3 & 38.5 / 88.5 / 92.3 & 46.2 / 88.5 / 92.3 & \textbf{26.9 / 88.5 / 92.3} \\
\textit{Gaussian} & 61.5 / 80.8 / 84.6 & 42.3 / 80.8 / 84.6 & 38.5 / 80.8 / 84.6 & 46.2 / 80.8 / 84.6 & \textbf{26.9 / 80.8 / 84.6} \\
\textit{Spatial} & 61.5 / 69.2 / 73.1 & 42.3 / 69.2 / 73.1 & 38.5 / 69.2 / 73.1 & 46.2 / 69.2 / 73.1 & \textbf{26.9 / 69.2 / 73.1} \\
\textit{Temporal} & 61.5 / 76.9 / 80.8 & 42.3 / 76.9 / 80.8 & 38.5 / 76.9 / 80.8 & 46.2 / 76.9 / 80.8 & \textbf{26.9 / 76.9 / 80.8} \\
\textit{Smoothing} & 61.5 / 73.1 / 76.9 & 42.3 / 73.1 / 76.9 & 38.5 / 73.1 / 76.9 & 46.2 / 73.1 / 76.9 & \textbf{26.9 / 73.1 / 76.9} \\
\textit{Body-LDP} & 61.5 / 65.4 / 69.2 & 42.3 / 65.4 / 69.2 & 38.5 / 65.4 / 69.2 & 46.2 / 65.4 / 69.2 & \textbf{26.9 / 65.4 / 69.2} \\
\textit{DMM} & 61.5 / 46.2 / 50.0 & 42.3 / 46.2 / 50.0 & 38.5 / 46.2 / 50.0 & 46.2 / 46.2 / 50.0 & \textbf{26.9 / 46.2 / 50.0} \\
\textit{DMM-Gaussian} & 61.5 / 42.3 / 46.2 & 42.3 / 42.3 / 46.2 & 38.5 / 42.3 / 46.2 & 46.2 / 42.3 / 46.2 & \textbf{26.9 / 42.3 / 46.2} \\
\textit{DMM-Spatial} & \textbf{61.5 / 30.8 / 38.5} & \textbf{42.3 / 30.8 / 34.6} & \textbf{38.5 / 30.8 / 34.6} & \textbf{46.2 / 30.8 / 38.5} & \textbf{\ul{26.9 / 30.8 / 30.8}} \\
\textit{DMM-Temporal} & 61.5 / 38.5 / 42.3 & 42.3 / 38.5 / 42.3 & 38.5 / 38.5 / 42.3 & 46.2 / 38.5 / 42.3 & \textbf{26.9 / 38.5 / 42.3} \\
\textit{DMM-Smoothing} & 61.5 / 34.6 / 38.5 & 42.3 / 34.6 / 38.5 & 38.5 / 30.8 / 38.5 & 46.2 / 34.6 / 38.5 & \textbf{26.9 / 34.6 / 38.5} \\

\bottomrule
\end{tabular}
}
\label{tab:rounded_gaze_body_multireid}
\end{table*}

 \textbf{Note on Eye Privacy Results.}  
As discussed in Section~5.2, we adopt  {privacy configuration for eye telemetry} identified by Wilson et al.~\cite{wilson2024PrivacyPreserving} as optimal for preserving utility in competitive XR tasks. Therefore, we do not re-evaluate trade-offs for eye  {telemetry} in this setting and instead focus our results on body and multimodal privacy mechanisms.

Considering the results discussed in this section, we arrive at the following answer to RQ2:

\textit{In competitive XR settings, \textit{Smoothing} remains the most effective eye privacy mechanism, as established by prior work~\cite{wilson2024PrivacyPreserving}. For body privacy mechanisms, our results show that \textit{DMM-Spatial} offers the best trade-off, significantly reducing re-identification risk while preserving high in-game performance.
}

\section{Evaluation of Anonymization in Multimodal XR}

This section investigates re-identification risks associated with multimodal XR telemetry, extending prior unimodal analyses to both casual and competitive settings and addresses the following research question:

\textit{
    RQ3: When streaming both eye and body  {telemetry} data, how well do  {paired} multimodal privacy mechanisms protect against re-identification in casual and competitive XR applications, while maintaining real-time usability?
}

To maintain consistency with the unimodal analyses, we reuse the OpenNEEDS and Chimera datasets for multimodal evaluation. OpenNEEDS, which includes eye and body telemetry from low-intensity tasks such as reading and drawing, represents casual XR applications where spatial precision is prioritized over speed. Chimera, constructed by pairing gaze from GazeBaseVR with body motion from BeatLeader using an established chimeric fusion method, models the long-term drift regime in competitive contexts. In this regime, high-intensity movements remain within each modality, but fine-grained cross-modal coordination has eroded, reflecting the toughest-case scenario in archived telemetry.

We also incorporate the dataset by Wilson et al.~\cite{wilson2024PrivacyPreserving}, used in prior XR privacy research, which contains naturally synchronized eye and body telemetry from 26 participants performing an XR cooking task involving rapid object manipulation, navigation, and visual search. This elicits frequent gaze shifts and active body movements, producing tightly correlated multimodal signals characteristic of competitive XR in short-term, high-interactivity sessions. Together, Chimera and Wilson span the full multimodal correlation spectrum, enabling privacy mechanism evaluation under both the hardest long-term drift and the most synchronized short-term conditions.

% % \textbf{Utility Metrics.}  
% The utility metrics applied in this multimodal analysis remain consistent with those used in the unimodal setting, allowing for direct comparisons. In casual XR applications, angular distance is used to assess the accuracy of gaze-based interactions, while Euclidean distance quantifies the spatial distortion introduced to body movements. These metrics are particularly relevant in precision-critical tasks such as object selection and hand–eye coordination, where even small perturbations may affect interaction fidelity.

% In competitive XR scenarios, user performance is evaluated using normalized Beat Saber score for body motion and angular distance for eye movement. These metrics capture the distinct contributions of each modality: body motion influences reaction speed and target alignment, while gaze direction affects targeting accuracy. Retaining these task-specific utility measures allows us to evaluate whether coordinated multimodal privacy mechanisms can preserve interaction performance in real-time, high-pressure environments.

 \textbf{Experimental Setup.}  
Our multimodal re-identification experiments follow the same 4-fold cross-validation protocol described in earlier sections, using EKYT~\cite{lohr2022Eyeb, aziz2024exploiting} as the identification model. The input representation  {consists of} both gaze and body movement features. Since parameter sweeps were already conducted in the unimodal analyses in Sections 4 and 5 \textcolor{black}{for each application type}, we do not repeat them here. Instead, we select discrete points from those trade-off curves that satisfy the utility thresholds outlined below. This approach ensures that the chosen privacy mechanisms align with task-relevant usability requirements for real-time XR applications.

\begin{itemize}
  \setlength\itemsep{0em}
    \item \textbf{Eye movements:} In casual XR  {applications}, parameters are selected to constrain angular deviation to $\leq 2^\circ$, a threshold shown to preserve interaction accuracy in spatially precise gaze tasks~\cite{schuetz2020psychophysics}. For competitive XR, we adopt the configuration identified by Wilson et al.~\cite{wilson2024PrivacyPreserving}, which maintains user performance based on Area of Interest (AOI) accuracy in high-speed gaze tasks.

    \item \textbf{Body movements:} For casual applications, we enforce a Euclidean deviation threshold of 21 cm, which has been shown to preserve spatial task fidelity in XR environments~\cite{wentzel2020improving}. For competitive XR, we apply a minimum utility threshold of $80\%$ of the original Beat Saber score, reflecting acceptable performance degradation for high-precision gameplay tasks~\cite{nair2023deep}.
\end{itemize}

 \textbf{Identification Protocol.}  
We adopt a two-session matching protocol to evaluate re-identification risk. For each participant, embeddings from the first session are used as the gallery and those from the second session as the probe. Cosine similarity is computed between probe–gallery pairs, and identification performance is reported as Rank-1 IR. Privacy mechanisms are applied to both gallery and probe samples, simulating a strong adversary capable of collecting privatized data or applying consistent transformations at inference time. Dataset splits follow the same protocol as in earlier sections. For OpenNEEDS \textcolor{black}{and Wilson et al.}, we use 4-fold cross-validation, with each fold comprising 75\% of participants for training and 25\% for testing. For Chimera, we reserve 46 participants for testing, and use the remaining 361 participants for training and validation.

\subsection{Results}

Tables~\ref{tab:gaze_body_multireid} and~\ref{tab:rounded_gaze_body_multireid} present Rank-1 IR for gaze-only, body-only, and  {} multimodal configurations in the OpenNEEDS (casual) and Chimera \& Wilson et al. (competitive) datasets, respectively. Each table reports identification rates under various combinations of eye and body privacy mechanisms.

 \textbf{Casual XR (OpenNEEDS).}  
In the absence of any privacy protection, multimodal Rank-1 IR reaches 80.3\%, underscoring the substantial identity leakage when eye and body signals are streamed together. Applying eye-only mechanisms reduces this risk modestly (e.g., smoothing eye  {telemetry} reduces multimodal IR to 72.9\%), suggesting that eye  {telemetry} contributes less than body data to multimodal identification.

In contrast, privacy applied to body  {telemetry} has a much greater effect. For instance, DMM-Smoothing on body  {telemetry} alone reduces multimodal IR to 33.6\%, even when gaze remains unprotected. The best result is achieved by  {pairing} DMM-Smoothing for body  {telemetry} with Smoothing for gaze, reducing multimodal IR to 26.3\%. This combination also lowers body and gaze IRs to 21.1\% and 27.6\%, respectively, indicating strong privacy protection across modalities.

 \textbf{Competitive XR (Chimera \textcolor{black}{and Wilson et al.}).}  
Re-identification risks are even more pronounced in performance-driven settings. For example, in Chimera dataset, without any protection, multimodal Rank-1 IR reaches 84.8\%, the highest observed in our experiments. Again, applying eye-only mechanisms such as smoothing lowers gaze IR (to 23.9\%) but has minimal impact on multimodal risk, which remains above 80\% when body movements are unprotected.

The most effective protection is achieved using DMM-Spatial for body  {telemetry}  {paired} with Smoothing for gaze  {telemetry}, yielding the lowest multimodal IR of 26.1\%. This configuration also achieves the lowest body IR (26.1\%) while holding gaze IR constant at 23.9\%. Other DMM-based combinations (e.g., DMM-Temporal and DMM-Smoothing) also show strong reductions, but DMM-Spatial consistently outperforms them in the competitive setting. \textcolor{black}{A similar trend appears in the real-world Wilson et al. dataset: eye-tracking protections alone, such as Smoothing, provide little reduction in multimodal risk, whereas combining Smoothing with DMM-Spatial for body telemetry again yields the strongest overall defense. The agreement in top-ranked pairings across both the chimera and natural datasets supports the robustness of our competitive findings. }

\ethanwilsonroundone{It stands out to summarize the results when we're still midway through the results section (Section 6).  This might flow better if moved to the start of the discussion.}
\aibragimovroundone{Agree. I removed the summary section from here as this is repeated in the discussion section.}

\review{

\textbf{Summary.}  
Across both application types, multimodal re-identification risks remain high when either modality is left unprotected. Eye-only and body-only mechanisms offer limited benefit in isolation, while paired body and eye protection-particularly DMM variants for body  telemetry and smoothing for gaze  telemetry, consistently produce the best trade-offs. Notably, the optimal mechanism pairing differs by context: DMM-Smoothing performs best in casual settings, while DMM-Spatial is most effective in competitive applications. These results reinforce the importance of task-aware, multimodal privacy strategies in extended reality systems.  Additionally, while prior work has shown that residual identity information can persist even when unimodal signals are independently privatized, our results suggest that well-paired mechanisms across modalities can suppress such leakage. In particular, we observe minimal performance gain from multimodal over unimodal re-identification in these cases, indicating effective mitigation of cross-modal identity cues.}{}

 {

Considering the results discussed in this section, we arrive at the following answer to RQ3: 

\textit{
Multimodal re-identification risk remains high when only one modality is protected. Well-paired mechanisms that anonymize both gaze and body movements are necessary to achieve meaningful privacy protection. In casual XR settings, the combination of DMM-Smoothing applied to body telemetry and Smoothing applied to eye telemetry achieves the best trade-off, reducing multimodal Rank-1 identification rate to 26.3\% while preserving interaction fidelity. In competitive scenarios, the most effective configuration uses DMM-Spatial for body telemetry and Smoothing for gaze telemetry, reducing the identification rate to 26.1\%.
}

\section{Open Science Policy: XR Privacy SDK}
\subsection{XR Privacy SDK}

\textcolor{black}{To support reproducibility and promote the broader adoption of privacy-preserving techniques in immersive environments, we introduce the \texttt{XR-Privacy} software development kit (SDK). This SDK is compatible with Unity and can be easily integrated into existing XR applications. Its primary goal is to provide developers with simple, configurable tools for protecting user identity via real-time eye and body movement anonymization.}

% \textcolor{black}{The SDK works by allowing users and developers to apply appropriate privacy levels for selected application type. It applies customizable privacy mechanisms to reduce re-identification risks.}

\begin{figure}[h]
    \centering
    \includegraphics[width=0.31\textwidth]{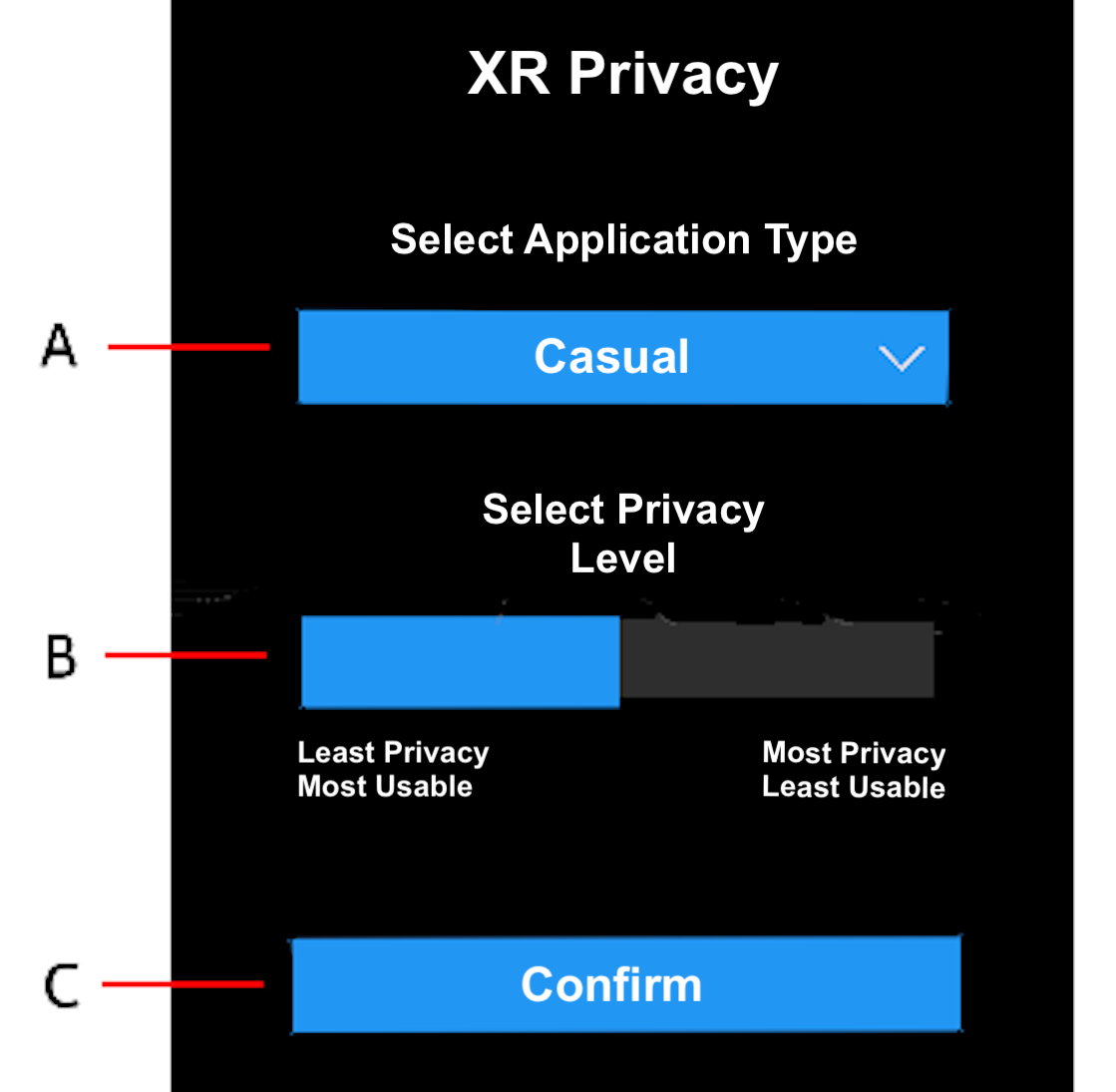}
    \caption{GUI of XR Privacy. (A) Dropdown to select application type. (B) Slider to configure privacy level based on the tradeoff between privacy-utility. (C) Button to confirm selected settings.}
    \label{fig:xr-ui}
\end{figure}

\textcolor{black}{Figure~\ref{fig:xr-ui} shows the graphical user interface of XR-Privacy SDK, which includes three core components. The application type selector (A) allows users to specify the general purpose of the experience (casual or competitive), which determines the appropriate privacy mechanisms and utility constraints. The privacy slider (B) allows users or developers to choose a position along the privacy–usability spectrum, with options ranging from ``Least Privacy, Most Usable'' to ``Most Privacy, Least Usable''. By default, the slider endpoints are mapped to the empirical bounds observed in our privacy-utility trade-off analyses (Figures 2 and 3): the leftmost position corresponds to the minimum privacy levels that still provide meaningful protection, while the rightmost position corresponds to the maximum privacy levels before violating application-specific usability thresholds established in Sections 4-6. Intermediate slider positions interpolate between these empirically-determined bounds according to the optimal configurations identified for each application type. Developers retain full flexibility to override these default mappings and define custom privacy-utility ranges based on their specific application requirements. Once configured, the ``Confirm'' button (C) initializes the corresponding privacy mechanism pairing.}

\begin{figure}[h]
    \centering
    \includegraphics[width=\linewidth]{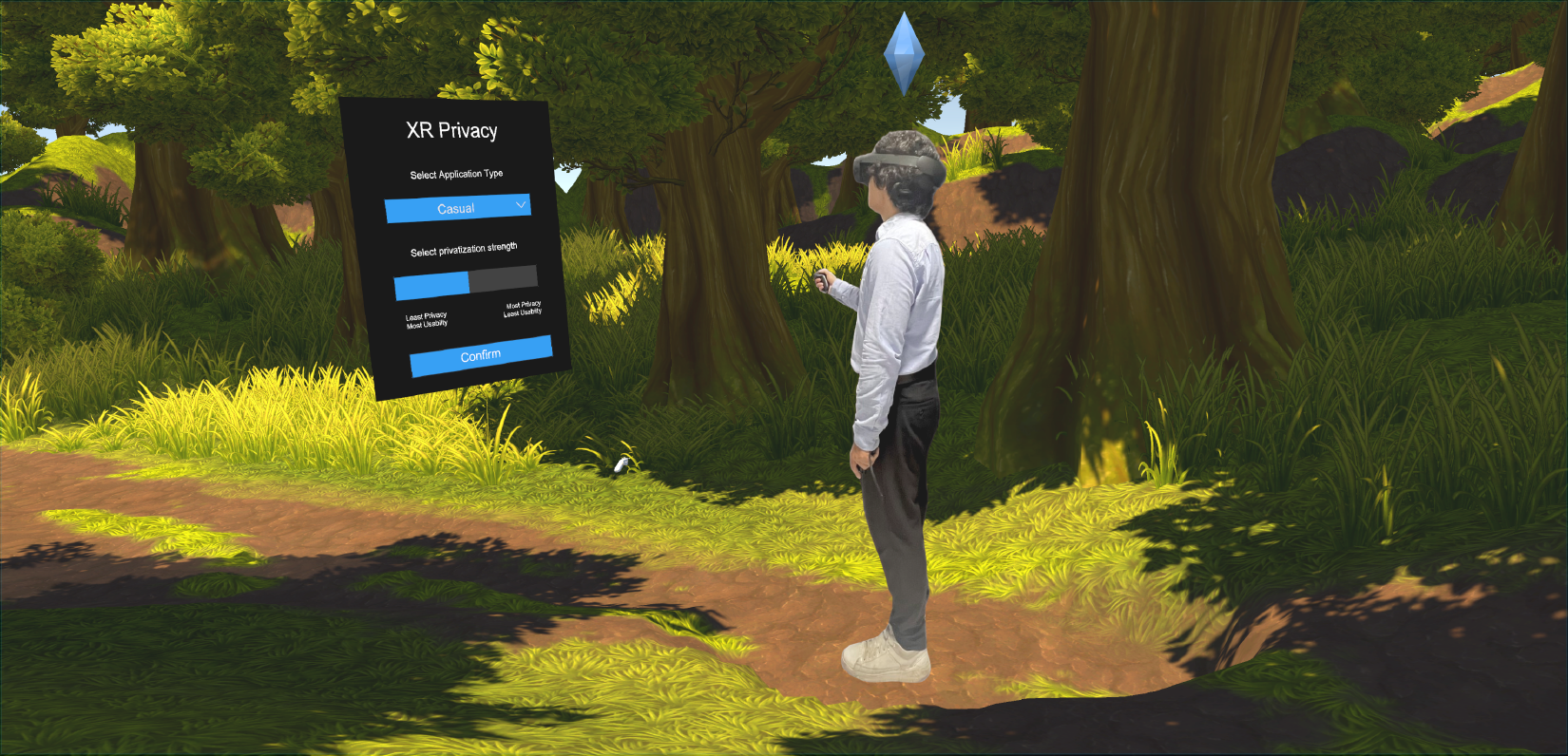}
    \caption{Mixed reality photo of a player using XR-Privacy SDK within a custom Unity-based virtual environment.}
    \label{fig:xr-use}
\end{figure}

\textcolor{black}{As shown in Figure~\ref{fig:xr-use}, the SDK is designed to be easily deployable inside existing Unity scenes. It presents the privacy configuration panel in 3D space, enabling users to interact with it using XR controllers or gaze tracking. This interface is rendered as part of the virtual scene, ensuring that users can make informed decisions about data sharing before any sensitive motion data is accessed.}

To demonstrate real-world impact, we integrated our SDK into multiple Unity Microgames, including the LEGO Microgame~\cite{lego_microgame}, Karting Microgame~\cite{unity_karting} and FPS Microgame~\cite{unity_fps}. Integration required less than 15 minutes in each case, with approximately six additional lines of code to configure privacy parameters. Across all projects, runtime performance showed negligible overhead: frame rates decreased only slightly (37.61 → 37.45 FPS) after enabling privacy mechanisms. These results highlight that applications built from widely adopted Unity templates can be readily extended with our SDK, underscoring its practical deployability in real development workflows.

\textcolor{black}{The SDK can be downloaded via the Unity Package Manager or installed manually using the following repository link:}

\url{https://anonymous.4open.science/r/XR-Privacy-SDK-86FF}

\section{Discussion \& Conclusion}

 \textbf{Timeline of XR Privacy Research and Our Contributions.} Prior work has progressively addressed XR privacy threats, but significant gaps remain. For gaze telemetry, Wilson et al.~\cite{wilson2024PrivacyPreserving} achieved 14.1\% re-identification rates using smoothing. For body telemetry, Nair et al.~\cite{nair2023deep} developed DMM, reducing re-identification to 3.7\% after showing that earlier approaches like Body-LDP failed against stronger adversaries.

Recent work by Aziz et al. established critical vulnerabilities in multimodal XR privacy, demonstrating that unimodal protections are insufficient and that residual identity information persists even when both modalities are privatized. Their evaluation of gaze smoothing with Body-LDP only reduced re-identification to 40.8\%. As the authors acknowledged, their study was limited to narrow datasets and mechanisms, lacking usability evaluations.

Our work addresses these gaps through: (1) large-scale evaluation of 14 mechanisms across 3 datasets, (2) a context-aware framework with application-specific utility thresholds, (3) improved pairings reducing re-identification to ~26\% while maintaining usability, and (4) an open-source XR-Privacy SDK for practical deployment.

\textcolor{black}{\textbf{Limitations \& Future Work.}} 
This study deliberately focused on re-identification to establish a rigorous foundation for multimodal XR privacy evaluation, as it represents the most widely recognized and benchmarked threat with established adversary models~\cite{aziz2024exploiting}. Future research can extend this framework to systematically investigate broader inference risks, including cognitive and demographic attribute detection such as mental health conditions~\cite{armstrong2012eye, giaretta2024security}, gender~\cite{sam2017Gender, jarin2023behavr}, and age~\cite{zhang2018old, giaretta2024security}.  

\textcolor{black}{We intentionally bounded our analysis to eye and body telemetry, since these essential XR control modalities cannot be disabled without eliminating control functionality. This ensures our findings address unavoidable privacy risks. Building on this foundation, future work can investigate privacy mechanisms for emerging optional modalities such as voice and facial expressions, which present unique opportunities for adaptive privacy protection that responds to user preferences and application requirements while preserving rich communication capabilities.}

\acknowledgments{
This work was supported in part by a Meta Privacy Enhancing Technology Award and from the National Science Foundation under grant CNS-2206950.}

\bibliographystyle{abbrv-doi}

\bibliography{template}

\end{document}